\newcommand{\be}[0]{\begin{equation}}
\newcommand{\ee}[0]{\end{equation}}
\newcommand{\ba}[0]{\begin{eqnarray}}
\newcommand{\ea}[0]{\end{eqnarray}}

\documentclass[published]{JHEP3} 

\JHEP{10(2004)062}

\usepackage{epsfig,multicol}

\newcommand\fverb{\setbox\pippobox=\hbox\bgroup\verb}
\newcommand\fverbdo{\egroup\medskip\noindent%
            \fbox{\unhbox\pippobox}\ }
\newcommand\fverbit{\egroup\item[\fbox{\unhbox\pippobox}]}
\newbox\pippobox

\title{Next-to-Leading order approximation of polarized valon and parton distributions
}

\author{ Ali
N. Khorramian $^{(a,d)}$, A. Mirjalili $^{(b,d)}$, and S. Atashbar
Tehrani $^{(c,d)}$

\\
\small{\it
$^{(a)}$ Physics Department, Semnan University, Semnan, Iran \\
$^{(b)}$ Physics Department, Yazd University, Yazd, Iran \\
$^{(c)}$ Physics Department, Persian Gulf University, Boushehr, Iran\\
$^{(d)}$ Institute for Studies in Theoretical Physics and
Mathematics (IPM), \\
\hspace{.48 cm}P.O.Box 19395-5531, Tehran, Iran \\

 }
\\

    E-mail:  \email{Alinaghi.Khorramian@cern.ch}, \email{mirjalili@ipm.ir}, \email{atashbar@ipm.ir}}
\received{August 26, 2004}        
\revised{October 26, 2004}
\accepted{October 26, 2004}        

\abstract{Polarized parton distributions and structure functions
of the nucleon are analyzed in the improved valon model. The valon
representation provides a model to represent hadrons in terms of
quarks, providing a unified description of bound state and
scattering properties of hadrons. Polarized valon distributions
are seen to play an important role in describing the spin
dependence of parton distributions in the leading order (LO) and
next-to-leading order (NLO) approximations. In the polarized case,
a convolution integral is derived in the framework of the valon
model. The Polarized valon distribution in a proton and the
polarized parton distributions inside the valon are necessary to
obtain the polarized parton distributions in a proton. Bernstein
polynomial averages are used to extract the unknown parameters of
the polarized valon distributions by fitting to the available
experimental data. The predictions for the NLO calculations of the
polarized parton distributions and proton structure functions are
compared with the LO approximation. It is shown that the results
of the calculations for the proton structure function, $xg_1^p$,
and its first moment, $\Gamma _{1}^p$, are in good agreement with
the experimental data for a range of values of $Q^{2}$. Finally
the spin contribution of the valons to the proton is calculated.}


\begin{document}
\section{Introduction}
Determination of  parton distributions in a nucleon in the
framework of quantum chromodynamics (QCD)  always involves some
model-dependent procedure. Instead of relying on mathematical
simplicity as a guide, we take a viewpoint in which  the physical
picture of the nucleon structure is emphasized. That is, we
consider the model for the nucleon which is compatible with the
description of the bound state problem in terms of three
constituent quarks. We adopt the view that these
 constituent quarks in the scattering problems should be regarded as the valence quark
clusters rather than point-like objects. They have been referred
to as $\it{valons}$. In the valon model, the proton consists of
two ``up'' and one ``down'' valons. These valons thus carry the
quantum numbers of the respective valence quarks. Hwa \cite{1}
found evidence for the valons in the deep inelastic neutrino
scattering data, suggested their existence and applied it to a
variety of phenomena. Hwa \cite{2} has also successfully
formulated a treatment of the low-$p_{T}$ reactions based on a
structural analysis of the valons. Some papers can  be found
 in which the valon model has been used to extract new
information for  parton distributions and  hadron structure functions \cite{3}.\\

Hwa and Yang in two papers \cite{4} refined the idea of the valon
model and extracted new results for the valon distributions. In a
recent paper, unpolarized parton distributions and hadronic
structure functions in the NLO approximation were extracted
\cite{5}.
The pion and meson structure functions in the valon
model framework has also been analyzed \cite{6, 7}.
In this paper we extend the idea of the valon model to both
polarized and unpolarized cases in the LO and NLO approximations.\\

The plan of the paper is as follows. In Subsec. 2.1, unpolarized
valon distributions are explained. In Subsec. 2.2 we define a
convolution integral, using a linear combination of spin-up and
down quark distributions. In Subsec. 2.3  the relationship between
unpolarized and polarized valon distributions in a proton is dealt
with. The discussion of the polarized valons for the non-singlet
and singlet cases is further refined in Subsec. 2.4.  $\cal {W}$
functions which link the polarized valon distributions and the
unpolarized ones are described in Subsec. 2.5. These contain some
unknown parameters which can be obtained  by fitting to the
experimental data. In Sec. 3 the moments of the polarized valon
and polarized parton distributions are analyzed in the NLO
approximation. Subsec. 4.1 describes the method of Ref.\cite{8} in
which one constructs averages of the measured polarized structure
functions weighted by suitably chosen Bernstein polynomials. These
experimental Bernstein averages are then fitted in Subsec. 4.2,
using the CERN subroutine MINUIT, to the QCD predictions for the
corresponding linear combinations of moments. In section 5 the
inverse Mellin transformation and the convolution integral are
used to calculate the polarized parton distributions in the
x-space. In Subsec. 6.1
 the polarized structure function, $xg_{1}$, for the
proton is calculated in the LO and NLO approximations, and
compared with the experimental data and some other theoretical
models. Our prediction for the first moment of proton structure
functions are presented  in this subsection for some values of
$Q^2$. The spin contribution of valons to the proton are discussed
and calculated in Subsec. 6.2. We also give our conclusions in
Sec. 7.

\section{ Valon distributions in unpolarized and polarized cases}

\subsection{Valon distributions}
The subject of nucleon structure has mainly been investigated from
two opposite viewpoints. On the one hand, one studies the bound
state in terms of three quarks, such as in a bag model, and
obtains various static properties of the nucleon and its
spectroscopic partners. On the other hand, one probes the nucleon
with high energy leptons in the hope of learning something about
its three constituent quarks but  finds structure functions that
can only be understood in terms of an infinite number of partons
(i.e., quarks, antiquarks and gluons). The two views can be
reconciled if we recognize that the constituents in the quark
model for the static nucleon  are not the same objects as the
quarks in the parton model. Since the scales of spatial resolution
are different for the  two problems, we may regard the former as
clusters of the latter without contradicting either view. The
clusters have the quantum numbers of the valence quarks, and have
been called valons, for short. More  precisely, a valon can be
defined as a valence quark and associated sea quarks and gluons
which arise in the dressing processes of QCD. In a bound state
problem these processes are virtual and a good approximation for
the problem is to consider a valon as an integral unit whose
internal structure cannot be resolved. In a scattering situation,
on the other hand, the virtual partons inside a valon can be
excited and be put on the mass shell. It is therefore more
appropriate to think of a valon as a cluster of partons with some
momentum distribution. The proton, for example, has three valons
which interact with each other in a way that is characterized by
the valon wave function, while they  respond independently in an
inclusive hard collision with a $Q^{2}$
dependence that can be calculated in QCD at high $Q^{2}$.\\

 In the static problems there is a little difference between the usual
constituent quarks and the valons, since the point-like nature of
the constituent quarks is not a crucial aspect of the description,
and has been assumed mainly for simplicity. But, in the scattering
problems it is important to recognize that the valons, being
clusters of partons, can not easily undergo scattering as a whole.
The fact that the bound-state problem of the nucleon can be well
described by three constituent quarks implies that the spatial
extensions of the valons do not overlap appreciably. A physical
picture of the nucleon in terms of three valons is then quite
analogous to the usual picture of the deuteron in terms of
two nucleons.\\

According to this description of the valon model,
$F_{2}^{p}(x,Q^{2})$ as a structure function (e.g. $F_2$ or
$xF_3$) is related to a valon structure function
${\cal{F}}_{2}^{v}$ corresponding to a valon $v$, by smearing of
the valon momentum in the nucleon. On the other hand, the
structure function of a hadron is obtained by convolution of two
distributions: the valon distributions in the proton and the
structure function for each valon. In the unpolarized situation we
may write:
\begin{equation}
F_{2}^{p}(x,Q^{2})=\sum_{v}\int_{x}^{1} dy G_{v/p}(y)
{\cal{F}}_{2}^{v}(\frac{x}{y},Q^{2})\;,
\end{equation}
where the summation is over the three valons. Here  $G_{v/p}(y)$
indicates the probability for the $v$-valon to have momentum
fraction $y$ in the proton. We shall also assume that the  three
valons carry all the momentum of the proton. This assumption is
reasonable provided that the exchange of  very soft gluons is
responsible for the binding. Eq.(2.1)  involves also the
assumption that in the deep inelastic scattering at high $Q^2$ the
valons are independently probed, since the shortness of
interaction time makes it reasonable to ignore the  response of
the spectator valons. Thus, through  Eq.(2.1) we have  broken up
the hadron structure problem into two parts. One part represented
by $G_{v/p}(y)$,  describes the wave functions of the proton in
the valon representation. It contains all the hadronic
complications due to the confinement. It is independent of $Q^2$
or the probe. The other part represented by
${\cal{F}}_{2}^{v}(\frac{x}{y},Q^{2})$, describes the virtual QCD
processes of the gluon emissions and quark-pair creation. It
refers to an individual valon independent of the other valons in
the proton and consequently also independent of the confinement
problem. It depends on $Q^2$ and
the nature of the probe.\\

Hwa and Zahir \cite{9} assumed a simple form for the exclusive
valon distribution which facilitated the phenomenological analysis
as follows
\begin{equation}
G_{UUD/p}(y_{1},y_{2},y_{3})=g (y_{1}y_{2})^{\alpha }y_{3}^{\beta
}\delta (y_{1}+y_{2}+y_{3}-1)\;,
\end{equation}
where $y_{i}$ is the momentum fraction of the $i$'th valon. The
$U$ and $D$ type inclusive valon distributions can be  obtained by
double integration over the unspecified variables
\begin{eqnarray}
G_{U/p}(y) &=&\int dy_{2}\int dy_{3}G_{UUD/p}(y,y_{2},y_{3}) \\
&=&g B(\alpha +1,\beta +1)y^{\alpha }(1-y)^{\alpha +\beta +1}\;,
\nonumber
\end{eqnarray}
\begin{eqnarray}
G_{D/p}(y) &=&\int dy_{1}\int dy_{2}G_{UUD/p}(y_{1},y_{2},y) \\
&=&g B(\alpha +1,\alpha +1)y^{\beta }(1-y)^{2\alpha +1}\;.
\nonumber
\end{eqnarray}
The normalization parameter $g$ has been fixed by \be
\int^{1}_{0}G_{U/p}(y)dy=\int^{1}_{0}G_{D/p}(y)dy=1\;, \ee \\
and is equal to $ g=[B(\alpha +1,\beta +1)B(\alpha +1,\alpha
+\beta +2)]^{-1}$,
where $B(m,n)$ is the Euler-Beta function.\\

R.C. Hwa and C. B. Yang \cite{4} have recalculated the unpolarized
valon distribution in the proton with a new set of parameters. The
new values of $\alpha $ , $\beta $ are found to be $\alpha=1.76 $
and
$\beta =1.05 $.\\

\subsection{Convolution integral}
To describe the quark distribution $q(x)$ in the valon model, one
can try to relate the polarized quark distribution functions
$q^\uparrow$ or $q^\downarrow$ to the corresponding valon
distributions $G^\uparrow$ and $G^\downarrow$. The polarized valon
can still have the valence and sea quarks that are polarized in
various directions, so long as the net polarization is that of the
valon. When we have only one distribution $q(x,Q^2)$ to analyze,
it is sensible to use the convolution in the valon model to
describe the proton structure in terms of the valons. In the case
that we have two quantities, unpolarized and polarized
distributions, there is a choice of which linear combination
exhibits more physical content. Therefore, in our calculations we
assume a linear combination of $G^\uparrow$ and $G^\downarrow$ to
determine respectively the unpolarized ($G$) and polarized
($\delta G$) valon
distributions . \\

To relate $q^\uparrow$ and $q^\downarrow$ to
both $G^\uparrow$ and $G^\downarrow$, we can consider linear
combinations as follows

\ba q^\uparrow_{i/p}(x,Q^{2})=\sum_{j}\int_{x}^{1}[
\alpha\;G^\uparrow_{j/p}(y)q^{\uparrow\uparrow}_{i/j}(\frac{x}{y},Q^{2})
+\beta\;G^\downarrow_{j/p}(y)q^{\downarrow\uparrow}_{i/j}(\frac{x}{y},Q^{2})]\frac{dy}{y}
\;, \ea and \ba q^\downarrow_{i/p}(x,Q^{2})=\sum_{j}\int_{x}^{1}[
\alpha'\;G^\uparrow_{j/p}(y)q^{\uparrow\downarrow}_{i/j}(\frac{x}{y},Q^{2})
+\beta'\;G^\downarrow_{j/p}(y)q^{\downarrow\downarrow}_{i/j}(\frac{x}{y},Q^{2})]\frac{dy}{y}
\;, \ea here $q^{\uparrow\uparrow}$ and $q^{\uparrow\downarrow}$
 respectively denote the probability of finding $q$-up  and
$q$-down in a $G$-up valon and  etc.\\

If we add and subtract the equations of (2.6, 2.7)
 we can determine the unpolarized and polarized
quark distributions as follows

\ba q_{i/p}(x,Q^{2})&\equiv&
q^\uparrow_{i/p}(x,Q^{2})+q^\downarrow_{i/p}(x,Q^{2}) \nonumber \\
&=&\sum_{j}\int_{x}^{1}( G^\uparrow_{j/p}[\alpha
q^{\uparrow\uparrow}_{i/j}+\alpha'q^{\uparrow\downarrow}_{i/j}]+
G^\downarrow_{j/p}[\beta
q^{\downarrow\uparrow}_{i/j}+\beta'q^{\downarrow\downarrow}_{i/j}])\frac{dy}{y};\ea
\ba \delta q_{i/p}(x,Q^{2})&\equiv&
q^\uparrow_{i/p}(x,Q^{2})-q^\downarrow_{i/p}(x,Q^{2}) \nonumber \\
&=&\sum_{j}\int_{x}^{1}( G^\uparrow_{j/p}[\alpha
q^{\uparrow\uparrow}_{i/j}-\alpha'q^{\uparrow\downarrow}_{i/j}]+
G^\downarrow_{j/p}[\beta
q^{\downarrow\uparrow}_{i/j}-\beta'q^{\downarrow\downarrow}_{i/j}])\frac{dy}{y}\;.\ea

Since it is reasonable to assume that
$q^{\uparrow\uparrow}=q^{\downarrow\downarrow}$ and
$q^{\uparrow\downarrow}=q^{\downarrow\uparrow}$, to obtain the
unpolarized and polarized quark distributions in the proton, we
need to choose in Eq.(2.8, 2.9) $\alpha=\alpha'=\beta=\beta'=1$.\\

Consequently we will have \ba
q_{i/p}(x,Q^{2})=\sum_{j}\int_{x}^{1}( G^\uparrow_{j/p}[q_{i/j}]+
G^\downarrow_{j/p}[q_{i/j}])\frac{dy}{y},\ea and \ba \delta
q_{i/p}(x,Q^{2})= \sum_{j}\int_{x}^{1}( G^\uparrow_{j/p}[\delta
q_{i/j}]- G^\downarrow_{j/p}[\delta q_{i/j}])\frac{dy}{y},\ea
where $q_{i/j}\equiv
q^{\uparrow\uparrow}_{i/j}+q^{\uparrow\downarrow}_{i/j}=
q^{\downarrow\uparrow}_{i/j}+q^{\downarrow\downarrow}_{i/j}$ and
$\delta q_{i/j}\equiv
q^{\uparrow\uparrow}_{i/j}-q^{\uparrow\downarrow}_{i/j}=
q^{\downarrow\downarrow}_{i/j}-q^{\downarrow\uparrow}_{i/j}$ are
representing unpolarized and polarized quark distributions in a
$j$-valon. Defining the unpolarized and polarized valon
distributions as \ba G_{j/p}\equiv\;G^\uparrow_{j/p}
+\;G^\downarrow_{j/p}\;, \ea and \ba \delta
G_{j/p}\equiv\;G^\uparrow_{j/p} -\;G^\downarrow_{j/p}\;, \ea we
will obtain \ba q_{i/p}(x,Q^{2})= \sum_{j}\int_{x}^{1}
q_{i/j}(\frac{x}{y},Q^{2})G_{j/p}(y)\frac{dy}{y}\;, \ea and
 \ba \delta
q_{i/p}(x,Q^{2})= \sum_{j}\int_{x}^{1}  \delta
q_{i/j}(\frac{x}{y},Q^{2})\delta G_{j/p}(y)\frac{dy}{y}\;. \ea As
we can see similarly  for the unpolarized case, the polarized
quark distribution can be related to a polarized valon
distribution.

\subsection{Spin dependence of the valon distributions}
The unpolarized  and polarized quark distributions are
respectively defined as
 \ba
 q(x)=q^{\uparrow}(x)+q^{\downarrow}(x)\;,\nonumber \ea
\ba
 \delta q(x)=q^{\uparrow}(x)-q^{\downarrow}(x)\;.\nonumber \ea
For the polarized parton distributions, $|\delta f(x,Q^2)|$, and
the unpolarized ones, $f(x,Q^2)$, positivity requirements at low
values of $Q^2$ imply the constraint \cite{10, 11} \be
 |\delta f(x,Q^2)|\leq
f(x,Q^2)\;, \ee where $f=u,\bar u, d, \bar d, s, \bar s, g$.
Furthermore, we have the following sum rules \cite{11} \be
 \Delta u+\Delta \bar u-\Delta d-\Delta
\bar d=A_3=1.2573\pm0.0028\;, \ee \be \Delta u+\Delta \bar
u+\Delta d+\Delta \bar d-2 (\Delta s+\Delta \bar
s)=A_8=0.579\pm0.025\;, \ee \be \Delta \Sigma =\sum_q (\Delta
q+\Delta \bar q)=A_8+3(\Delta s+\Delta \bar s)\equiv A_0\;, \ee
where with $n=1$, the first moment $\Delta f$ is defined by
 \be
 \Delta f(Q^2)=\int_{0}^{1}dx \delta
f(x,Q^2)\;. \ee\\

It is helpful to consider the above reasonable theoretical
constraints especially for the sea quark distributions. To
determine the polarized parton distribution functions (PPDF's),
the main step is to relate the polarized input densities to the
unpolarized ones \cite{11} using some intuitive theoretical
argument as the guideline.  We employ the general ansatz for the
polarized parton distributions of Ref.[11], and introduce the
following equations to relate the polarized and unpolarized valon
distributions
\begin{equation}
\delta G_{j/p}(y)=\delta {\cal {W}}_{j}(y)\times G_{j/p}(y)\;,
\end{equation}
here $j$ refers to $U$ and $D$ type valons. The functions $\delta
{\cal {W}}_{j}(y)$ play an essential role in constructing the
polarized valon distributions $\delta G_{j/p}(y)$ from unpolarized
ones, as we shall explain in Subsec. 2.5.\\

Let us define the Mellin moments of any polarized $U$ and
$D$-valon distributions in the proton, $\delta G_{U,D/p}(y)$, as
follows:

 \be \Delta M_{U/p}(n)\equiv\int_{0}^{1}y^{n-1} \delta
G_{U/p}(y) \,dy\;, \ee

 \be \Delta M_{D/p}(n)\equiv\int_{0}^{1}y^{n-1} \delta
G_{D/p}(y) \,dy\;. \ee

Using polarized valon distributions, the moments of polarized
valon distributions, $\Delta M_{U/p}(n)$ and $\Delta M_{D/p}(n)$,
can be easily calculated. The moments of polarized parton
distributions in the proton can be written like the unpolarized
case \cite{9} as follows
\begin{equation}
\Delta u_{v}(n,Q^{2})=2\Delta M^{NS}(n,Q^{2})\times\Delta
M_{U/p}(n)\;,
\end{equation}
\begin{equation}
\Delta {d_{v}}(n,Q^{2})=\Delta M^{NS}(n,Q^{2})\times\Delta
M_{D/p}(n)\;,
\end{equation}
\begin{equation}
\Delta {\Sigma }(n,Q^{2})=\Delta M^{S}(n,Q^{2})\times(2 \Delta
M_{U/p}(n)+\Delta M_{D/p}(n))\;.
\end{equation}
The factor 2 in Eqs.(2.24, 2.26) is due  to the existence of
two-$U$
type valons.\\

In the above equations, $\Delta M^{NS}(n,Q^{2})$ and $\Delta
M^{S}(n,Q^{2})$ are the moments of the non-singlet and singlet
sectors which describe parton distributions inside the valon. We
will introduce these functions in the LO and NLO approximations in
Sec. 3. As we will see, in the LO approximation the first moment,
$\Delta M^{NS}(n=1,Q^{2})$ and $\Delta M^{S}(n=1,Q^{2})$ will be
equal to $1$ for all values of $Q^2$. So, using Eqs.(2.24-2.26) we
have

\begin{equation}
\Delta u_v\equiv\Delta {u_{v}}(1,Q^{2})=2\Delta M_{U/p}(1)\;,
\end{equation}
\begin{equation}
\Delta d_v\equiv\Delta {d_{v}}(1,Q^{2})=\Delta M_{D/p}(1)\;,
\end{equation}
\begin{equation}
\Delta \Sigma \equiv\Delta {\Sigma }(1,Q^{2})=2 \Delta
M_{U/p}(1)+\Delta M_{D/p}(1)\;.
\end{equation}
It can be seen that in the LO approximation the first moment of
the polarized quark distributions is equal to the first moment of
the related polarized valon distributions. We should notice that
in the NLO approximation, the first moment of the polarized quark
distributions are $Q^2$-dependent, and the above equations are applicable  only for
$Q^2=Q^2_0$.\\

The $\Sigma$ symbol in Eq.(2.29) denotes $\sum_{q=u,d,s}(q+\bar
{q})$ and consequently we have the following definition of ${\Delta}{\Sigma}$

\be \Delta {\Sigma }(n,Q^{2})=\sum_{q=u,d,s}(\Delta
{q}(n,Q^2)+\Delta {\bar {q}}(n,Q^2))\;, \ee therefore the moment
of polarized sea quarks can be obtained as

\be \Delta {\bar {q}}(n,Q^2)=(\Delta {\Sigma }(n,Q^{2})-\Delta
{u_{v}}(n,Q^2)-\Delta {d_{v}}(n,Q^2))/2f\;, \ee where $f$ is the
number of active quark flavours. If we choose $n=1$ in Eq.(2.31)
and use Eqs.(2.27-2.29), we can see that for all $Q^2$ values in
the LO and for $Q^2=Q^2_0$ in the NLO approximations, there is no
contribution to the first moment of the sea quarks. As a result,
we can not precisely determine the spin components of the hadrons.
This difficulty is investigated in the following subsections.
\subsection{The improvement of polarized valons}

To resolve the above problem, first we improve the definition of
polarized valon distribution functions
 \ba \delta G_{j/p}
\rightarrow \left\{
\begin{array}{ll}
\delta {\cal {W}}^{'}_{j}(y)\times G_{j/p}(y)\  & {\rm for\;non-singlet\;case\;,} \\
\delta {\cal {W}}^{''}_{j}(y)\times G_{j/p}(y)\ & {\rm for\;
singlet\; case}\;.
\end{array}
 \right.
\ea Using the above equation and according to the Eqs.(2.27-2.29),
the first moment of the polarized $u$, $d$ and $\Sigma$
distribution functions can be  written as follows

 \be \Delta u_v=2\int_{0}^{1}dy\; [\delta {\cal {W}}^{'}_{U}(y)\times G_{U/p}(y)]\;,
 \ee
 \be \Delta
d_v=\int_{0}^{1}dy\; [\delta {\cal {W}}^{'}_{D}(y)\times
G_{D/p}(y)]\;, \ee \be \Delta \Sigma=\int_{0}^{1}dy\;
\left([2\;\delta {\cal {W}}^{''}_{U}(y)\times G_{U/p}(y)]+
[\delta {\cal {W}}^{''}_{D}(y)\times G_{U/p}(y)]\right)\;. \ee\\

The above equations can help us to consider the constraint of
Eqs.(2.17-2.19) for the improved polarized valon model with an
SU(3) flavour symmetry assumption. These constraints play the same
role as in the case of the unpolarized ones, Eq.(2.5), in
controlling the parameter values which will appear in the
polarized valon distributions.

\subsection{Description of the $\bf \cal {W} $ function }

To define the actual $y$-dependence of the ${\cal {W}}^{'}$ function, we parameterize  this function
as
\begin{equation}
\delta {\cal {W}}^{'}_{j}(y)=N_{j}y^{\alpha _{j}}(1-y)^{\beta
_{j}}(1+\gamma _{j}y+\eta _{j}y^{0.5})\;.
\end{equation}
As before the subscript $j$ refers to $U$ and $D$-valons.  \\

The motivation for choosing this functional form is that the term
$y^{\alpha _{j}}$ controls the low-$y$ behavior valon densities,
and $(1-y)^{\beta _{j}}$ that at large values $y$. The remaining
polynomial factor accounts for the additional medium-$y$ values.\\

For $\delta {\cal {W}}^{''}_{j}(y)$ we choose the following form
\begin{equation}
\delta {\cal {W}}^{''}_{j}(y)=\delta {\cal {W}}^{'}_{j}(y)\times
\sum_ {m=0}^{5} {\cal A}_{m}y^{\frac {m-1}{2}}\;.
\end{equation}\\
The additional term in the above equation, ($\sum$ term), serves
to control the behavior of the singlet sector at very low-$y$
values in such a way that we can extract the sea quark
contributions. Moreover, the functional form for $\delta {\cal
{W}}^{'}$ and $\delta {\cal {W}}^{''}$ give us the best fitting
$\chi^2$ value, as will be described in Sec.4. In these functions,
all of the parameters are unknown. Using the experimental data for
$g_1^p$ \cite{12, 13} and the Bernstein polynomials, we can fit
for the unknown parameters of Eqs.(2.36, 2.37).
\section{The NLO moments of PPDF's and the  structure function}
In this section  moments of the polarized parton distribution
functions (PPDF's) in the valon  are introduced. To calculate the
 $Q^2$ evolution of the polarized parton distributions, $\delta
f(x,Q^{2})$, we need to Mellin transform these functions, i.e.

\begin{equation}
\delta f^n(Q^2)=\int^1_0 x^{n-1} \delta f(x,Q^2)dx\;.
\end{equation}
The evolution of anomalous dimensions can be obtained by Mellin
transformation of splitting functions, and admit an expansion in
powers of the running coupling constant $\alpha_{s}(Q^{2})$

\begin{equation}
\delta d_{NS\pm}= \frac{\alpha_{s}(Q^{2})}{2\pi} \delta
d^{(0)}_{qq}(x) + \left(\frac{\alpha_{s}(Q^{2})}{2\pi}\right)^2
\delta d^{(1)}_{NS\pm} (x)+ ...\;,
\end{equation}
whose detailed $n$-dependence has been  specified in Ref.[10].
Here $\delta d^{(0)}_{qq}$ is unique whereas all higher anomalous
dimensions are scheme-dependent.\\

The leading and next-to-leading order solutions of the
renormalization group equation for the polarized moments can be
expressed entirely in terms of the evolution parameter
\begin{equation}
L(Q^{2})\equiv \frac{\alpha_{s} (Q^{2})}{\alpha_{s}
(Q^{2}_{0})}\;.
\end{equation}
The non-singlet (NS) moments evolve according to

\begin{equation}
\Delta M^{NS\pm}(n,Q^{2})=\left(
1-\frac{\alpha_{s}(Q^{2})-\alpha_{s} (Q^{2}_{0})}{2\pi }( \delta
d_{NS\pm}^{(1)n}-\frac{2\pi b^{\prime }}{b}\delta d_{qq}^{(0)n})
\right) L^{\delta d_{qq}^{(0)n}}\;,
\end{equation}
and the NLO running coupling constant is given by
\begin{equation}
\alpha_{s}(Q^{2})\cong\frac{1}{b\log \frac{Q^{2}}{\Lambda_{\overline{MS}} ^{2}} }-\frac{%
b^{\prime }}{b^3}\frac{\ln \left( \ln  \frac{Q^{2}}{\Lambda_{\overline{MS}} ^{2}}%
 \right) }{\left( \ln \frac{Q^{2}}{\Lambda_{\overline{MS}} ^{2}} \right)
^{2}}\;,
\end{equation}
where $b=\frac{33-2f}{12\pi }$ and $b^{\prime
}=\frac{153-19f}{24\pi ^{2}}$. In the above equations, we choose
$Q_0=1\;GeV^2$ as a fixed parameter and $\Lambda$ is an unknown
parameter which can be obtained  by fitting
to experimental data.\\

The evolution in the flavor singlet and the gluon sector of the
moments are governed by a $2\times 2$  anomalous  dimension
matrix, with the following  explicit solution

\begin{equation}
\left(
\begin{array}{c}
\Delta M^{S }(n,Q^{2}) \\
\Delta M^{gq}(n,Q^{2})
\end{array}
\right) =\left( L^{\delta \hat{d}^{(0)n}}+\frac{\alpha_{s}(Q^{2})}{2\pi }\hat{U}L^{\delta \hat{d}^{(0)n}}-%
\frac{\alpha_{s}(Q^{2}_{0})}{2\pi }L^{\delta
\hat{d}^{(0)n}}\hat{U}\right) \left(
\begin{array}{c}
1 \\
1
\end{array}
\right)\;,
\end{equation}
 where $\Delta M^{gq}$ indicates the spin dependent quark-to-gluon
 evolution function. All the associated functions in the above equation have been defined in
 Ref.[14].\\

Using  Eqs.(3.4, 3.6), the moments of $NS$, $S$ and $gq$ as a
function of $n$ are calculable. These moments in the LO and
NLO approximations at $Q^2=3\; GeV^2$ are presented in Fig. 1. \\

Determination of the moments of parton distributions in a proton
can be done strictly through the moments of the polarized valon
distributions. The parton distributions in the x-space which will
be calculated, are $\delta u_v$, $\delta d_v$, $\delta \Sigma$ and
$\delta g$. The moments of these distributions are denoted
respectively by: $\Delta{u_{v}}(n,Q^{2})$,
$\Delta{d_{v}}(n,Q^{2})$, $\Delta{\Sigma }(n,Q^{2})$ and $\Delta
g(n,Q^{2})$. Therefore, the moments of the polarized $u$ and
$d$-valence quark in a proton are convolutions of two moments:
\begin{equation}
\Delta{u_{v}}(n,Q^{2})=2\Delta M^{NS}(n,Q^{2})\times\Delta
M^{'}_{U/p}(n)\;,
\end{equation}
\begin{equation}
\Delta{d_{v}}(n,Q^{2})=\Delta M^{NS}(n,Q^{2})\times\Delta M^{'}_
{D/p}(n)\;,
\end{equation}
The moment of the polarized singlet distribution ($\Sigma $) is as
follows:
\begin{equation}
\Delta{\Sigma }(n,Q^{2})=\Delta M^{S}(n,Q^{2})(2 \Delta
M^{''}_{U/p}(n)+\Delta M^{''}_{D/p}(n))\;.
\end{equation}

For the leading and next-to-leading terms in Eqs.(3.4, 3.6), for
the NS and S sectors, we can see that for all values of
 $Q^2$ in the LO and for $Q^2={Q_0}^2$ in the NLO, the first moments of $\Delta M^{NS}(1,Q^{2})$
 and $\Delta M^{S}(1,Q^{2})$ will be equal to one, and consequently the first moment of the
 polarized $u$ and $d$ quarks
are respectively proportional to the first moment of the polarized
$U$ and $D$
type valon distributions.\\

For the gluon distribution we have
\begin{equation}
\Delta g(n,Q^{2})=\Delta M^{gq}(n,Q^{2})(2 \Delta
M^{'}_{U/p}(n)+\Delta M^{'}_{D/p}(n))\;,
\end{equation}
where $\Delta M^{gq}(n,Q^{2})$ is the quark-to-gluon evolution
function, given in Eq.(3.6).\\

The NLO moment contribution of $g_1(x,Q^2)$ \cite{14} has the following form

\begin{equation}
g_1(n,Q^2)=\frac{1}{2}\sum\limits_q
e^2_q\{(1+\frac{\alpha_s}{2\pi}\delta C^n_q) [\Delta
q(n,Q^2)+\Delta\bar q(n,Q^2)] + \frac{\alpha_s}{2\pi}2\delta
C^n_g\Delta g(n,Q^2)\}\;,
\end{equation}
where $\Delta q(n,Q^2)=\Delta{{q_v(n,Q^2)}}+\Delta\bar q(n,Q^2)$,
$\Delta\bar q(n,Q^2)$ and $\Delta g (n,Q^2)$ are moments of the
polarized parton distributions in a proton. $\delta C^n_q$,
$\delta C^n_g$ are also the $n$-th moments of spin-dependent
Wilson coefficients given by
\begin{equation}
\delta C^n_q= \frac{4}{3}\biggl[-S_2(n)+(S_1(n))^2
+\left(\frac{3}{2}- \frac{1}{n(n+1)}\right) S_1(n) +\frac{1}{n^2}
+\frac{1}{2n}+\frac{1}{n+1}-\frac{9}{2} \biggr]\;,
\end{equation}
and
\begin{equation}
\delta C^n_g=\frac{1}{2}[-\frac{n-1}{n(n+1)}(S_1(n)+1)
-\frac{1}{n^2}+ \frac{2}{n(n+1)}] \;,
\end{equation}
with $S_{k}(n)$ defined as in Ref.[14].\\

So far the moments of the polarized parton distributions have been
determined and the moments of the polarized proton structure
function in the NLO approximation can be obtained by inserting the
required distribution functions in Eq.(3.11). There are some
unknown parameters in Eqs.(2.36, 2.37) and Eq.(3.5) which they
also appear in Eqs.(3.7-3.10). It is obvious that the final form
for $g_1(n,Q^2)$ involves the total of 17 unknown parameters. If
the parameters can be  obtained  then the computation of all
moments of the polarized parton distributions and the structure
function, $g_1(n,Q^2)$, are possible. \\

\section{QCD analysis method}

\subsection{The Bernstein averages of moments} In the phenomenological
investigations of the  structure functions, for a given value of
$Q^2$, only  a limited number of experimental points,
 covering a partial range of values of $x$, are available. Therefore,
one cannot directly determine the moments.
 A method devised to deal with this situation
 is to take  averages of the structure function weighted by suitable polynomials.
 We can compare theoretical predictions with the experimental results for the Bernstein
 averages, which are defined by \cite{15}
\be g_{n,k}(Q^2){\equiv}\int_{0}^{1}dxp_{nk}(x)g_1(x,Q^2)\;, \ee
where $p_{n,k}(x)$ are the Bernstein polynomials,
 \be
p_{n,k}(x)=\frac{\Gamma (n+2) }{\Gamma (k+1) \Gamma (n-k+1)
}x^k(1-x)^{n-k}\;,
 \ee
 and are normalized to unity,
$\int_{0}^{1}dxp_{n,k}(x)=1$. Therefore, the integral (4.1)
represents an average of the function $g_{1}(x, Q^2)$ in the
region
${\bar{x}}_{n,k}-\frac{1}{2}\Delta{x}_{n,k}{\leq}x{\leq}{\bar{x}}_{n,k}+\frac{1}{2}\Delta{x}_{n,k}$
where ${\bar{x}}_{n,k}$ is the average of $x$ which is very near
to the maximum of $p_{n,k}(x)$, and $\Delta{x}_{n,k}$ is the
spread of ${\bar{x}}_{n,k}$. The key point is that values of $g_1$
outside this interval have a small contribution to the above
integral, as $p_{n,k}(x)$ decreases to zero very quickly. In order
to ensure the equivalence of the integral (4.1) to the same
integral in the range
$x_1={\bar{x}}_{n,k}-\frac{1}{2}\Delta{x}_{n,k}$ to
$x_2={\bar{x}}_{n,k}+\frac{1}{2}\Delta{x}_{n,k}$, we have to use
the normalization factor of $\int_{x_1}^{x_2}dxp_{n,k}(x)$ in the
denominator of equation (4.1) which obviously is not equal to $1$.\\

By a  suitable choice of $n$, $k$ we manage to adjust the region
where the average is peaked to that in which we have experimental
data \cite{12, 13}.
   Using the binomial expansion in Eq.(4.2), it follows that the averages of $g_1$
   with $p_{n,k}(x)$ as weight functions, can be obtained in terms of odd and even moments,
\be
g_{n,k}=\frac{{(n-k)!}{\Gamma(n+2)}}{\Gamma(k+1)\Gamma(n-k+1)}\sum_{l=0}^{n-k}
\frac{(-1)^l}{l!(n-k-l)!}\int_{0}^{1}x^{\left((k+l+1)-1\right)}{g_1(x,Q^2)}dx\;,
\ee and  using the definition of Mellin moments of any hadron
structure function we have
 \be
g_{n,k}=\frac{{(n-k)!}{\Gamma(n+2)}}{\Gamma(k+1)\Gamma(n-k+1)}\sum_{l=0}^{n-k}
\frac{(-1)^l}{l!(n-k-l)!}\;{g_1({(k+l)+1}},Q^2)\;. \ee

We can only include a Bernstein average, $g_{n,k}$, if we have
experimental points covering the whole range
 [${\bar{x}}_{n,k}-\frac{1}{2}\Delta{x}_{n,k}, {\bar{x}}_{n,k}+\frac{1}{2}\Delta{x}_{n,k}$]
 \cite{16}. This means that with the available experimental data we can only use
 the following 41 averages,\\

\begin{center}
$g_{2,1}^{(\exp )}(Q^{2}),g_{2,2}^{(\exp )}(Q^{2})$, ...,
$g_{13,10}^{(\exp )}(Q^{2})$.
\end{center}

Another restriction which we assume here, is  to ignore the
effects of moments with high order $n$ which do not strongly
constrain the fits. To obtain these experimental averages from the
E143 and SMC data \cite{12, 13}, we fit $x{g_1}(x,{Q^2})$ for each
bin in ${Q}^{2}$ separately, to the convenient phenomenological
expression \be
{xg_{1}}^{\hspace{-.12cm}{(phen)}}={\cal{A}}x^{\cal{B}}(1-x)^{\cal{C}}\;.
\ee This form ensures zero values for ${xg_{1}}$ at $x=0$, and
$x=1$. In Table I we have presented the numerical values of
$\cal{A},\cal{B}$ and $\cal{C}$ for $Q^2=3, 5, 10\;GeV^2$. A
theoretical justification of Eq.(4.5) may be found in
Ref.\cite{17}. Using Eq.(4.5) with the fitted values of
${\cal{A}},{\cal{B}},{\cal{C}}$ one can then compute
${g}_{n,k}^{(exp)}({Q}^{2})$ using Eq.(4.3), in terms of Gamma
functions. Some sample experimental Bernstein averages are plotted
in Fig. 2 in the LO and NLO approximations. The errors in the
${g}_{n,k}^{(exp)}(Q^2)$ correspond to allowing the E143 and SMC
data for $x{g}_{1}$ to vary within the experimental error bars,
including the experimental systematic and statistical errors
\cite{12, 13}. We have only included data for
${Q}^{2}{\ge}3{\rm{GeV}}^{2}$, this has the merit of simplifying
the analysis by avoiding evolution through flavor thresholds.
\subsection{QCD fits to the Bernstein averages for $\bf g_1$}

Using Eq.(4.4), the 41 Bernstein averages ${g}_{n,k}({Q}^{2})$ can
be written in terms of odd and even moments $g_1(n,{Q}^{2})$: \ba
&&{g_{2,1}}(Q^2)=6\left(g_1(2,Q^2)-g_1(3,Q^2)\right)\;,
\nonumber\\
&&{g_{2,2}}(Q^2)=3\left(g_1(3,Q^2)\right)\;,
\nonumber\\
&&{g_{3,1}}(Q^2)=12\left(g_1(2,Q^2)-24g_1(3,Q^2)+12g_1(4,Q^2)\right)\;,
\nonumber\\
&&{g_{3,2}}(Q^2)=12\left(g_1(3,Q^2)-g_1(4,Q^2)\right)\;,
\nonumber\\
&& \vdots\nonumber\\
&&{g_{13,10}}(Q^2)=4004.0\left(g_1(11,Q^2)
-g_1(14,Q^2)\right)\;,\nonumber\\
&&-12012\left(g_1(12,Q^2)-g_1(13,Q^2)\right)\;. \ea\\
The unknown parameters according to Eqs.(2.36, 2.37) will be,
$N_U,\alpha_U,\beta_U,\;\cdots,N_D,\ \alpha_D\;$
\newline
$\;,\beta_D,\cdots,{\cal A}_0$, ${\cal A}_1,\cdots,{\cal A}_6$ and
finally $\Lambda_{\overline{MS}}$ from Eq.(3.5). Thus, there are
17 parameters to be simultaneously fitted to the experimental
${g}_{n,k}({Q}^{2})$ averages. Using the CERN subroutine MINUIT
\cite{18}, we defined a global ${\chi}^{2}$ for all the
experimental data points  and found an acceptable fit with minimum
${\chi}^{2}/{\rm{d.o.f.}}=1.26$ in the LO case and
${\chi}^{2}/{\rm{d.o.f.}}=0.94$ in the NLO case with the standard
error of order $10^{-3}$. The best fit is indicated by some sample
curves in Fig. 2. The minimum ${\chi}^{2}$ values for all 17
fitting parameters are listed in Table II.\\

 Now from Eqs.(2.36, 2.37), and using (2.32), we are able to determine
the improved polarized $U$ and $D$ valon distributions in the
proton. In Fig. 3 we have  plotted the LO and NLO approximation
results of $\sum _{j}^{}y\delta G_{j/p}$ for the  non-singlet and
singlet polarized valon distributions as a function of $y$. In
this summation $j$ runs over $U,U,D$ valons. Since the polarized
valon distributions are determined, we can obtain all  polarized
quark distributions in the proton using the convolution integral
of Eq.(2.15).

\section{Polarized parton distributions in the $\bf{x}$-space}

Polarized valon distributions in the proton have been calculated
in Subsec. 2.4 and Sec. 4. Using the polarized parton structure
functions in a valon, it is then possible to extract polarized
parton structure functions in a proton. To obtain the
$z$-dependence of parton
distributions from the $n-$%
dependent exact analytical solutions in the Mellin-moment space,
one has to perform a numerical integral in order to invert the
Mellin-transformation
\begin{equation}
\delta f^k(z,Q^{2})=\frac{1}{\pi }\int_{0}^{\infty }dw Im[e^{i\phi
}z^{-c-we^{i\phi }}\Delta M_k(n=c+we^{i\phi },Q^{2})]\;,
\end{equation}
where the contour of the integration lies to the right of all
singularities of $\Delta M_k(n=c+we^{i\phi },Q^{2})$ in the
complex $n$-plane. For all practical purposes one may choose
$c\simeq 1,\phi =135^{\circ }$ and an upper limit of integration,
for any $Q^{2}$ , of about $5+10/\ln z^{-1}$, instead of $\infty
$, which guarantees stable numerical results
\cite{19, 20}.\\

Inserting in Eq.(5.1) the three functions for $k=NS,\;S,\;gq$,
i.e. $\Delta M^{NS}(n,Q^{2})$, $\Delta M^{S}(n,Q^{2})$ and $\Delta
M^{gq}(n,Q^{2})$ separately, we can obtain all polarized parton
distribution functions inside the valon as a function of $z$, and
at a fixed $Q^2$ value. We denote them by $\delta
f^{k}(z=\frac{x}{y},Q^2)$. These functions are independent of the
type of valon but depend on $Q^2$. As was discussed in Subsec.
2.2, the relationship between the polarized parton distributions
of a proton and the polarized parton distributions in a valon can
be given by Eq.(2.15) and now by using this equation, we can get
the following expressions for the polarized parton distributions
in a proton:

\ba \delta u_v(x,Q^2)&=&2\int_x^1  \delta
f^{NS}(\frac{x}{y},Q^2)\;[\delta {\cal {W}}^{'}_{U}(y)\times G_{U/p}(y)]\;\frac{dy}{y}\;,\nonumber \\
\delta d_v(x,Q^2)&=&\int_x^1  \delta
f^{NS}(\frac{x}{y},Q^2)\;[\delta {\cal {W}}^{'}_{D}(y)\times G_{D/p}(y)]\;\frac{dy}{y}\;,\nonumber \\
\delta \Sigma(x,Q^2)&=&\int_x^1  \delta
f^{S}(\frac{x}{y},Q^2)\left(2\;[\delta {\cal
{W}}^{''}_{U}(y)\times G_{U/p}(y)]+
[\delta {\cal {W}}^{''}_{D}(y)\times G_{D/p}(y)]\right)\;\frac{dy}{y}\;,\nonumber \\
\delta g(x,Q^2)&=&\int_x^1  \delta
f^{gq}(\frac{x}{y},Q^2)\left(2\;[\delta {\cal
{W}}^{'}_{U}(y)\times G_{U/p}(y)]+ [\delta {\cal
{W}}^{'}_{D}(y)\times G_{D/p}(y)]\right)\;\frac{dy}{y}\;.\nonumber \\
\ea In Fig. 4  we have presented the polarized
 parton distributions in a proton at
  $Q^{2}=3\;GeV^{2}$. These distributions were calculated in the LO
 approximation and compared  with some theoretical
models [21-23]. In Fig. 5 we have presented the same distributions
but in the NLO approximation.
\section{Results}
\subsection{Polarized structure function}
In the last section  PPDF's were calculated and now they can be
used to extract the polarized structure
function $g_{1}^{p}$ in the LO and NLO approximations.\\

According to the quark model, in the LO approximation, $g_{1}^{p}$
can be written as a linear combination of $\delta q$ and $\delta
\overline{q}$ \cite{14, 24},

\begin{equation}
g_{1}^{p}(x,Q^{2})=\frac{1}{2}\sum_{q} e_{q}^{2}[\delta
q(x,Q^{2})+\delta \overline{q}(x,Q^{2})]\;,
\end{equation}
where $e_{q}$ are the electric charges of the (light)
quark-flavours $q=u,d,s$. Here the sum usually runs over the light
quark-flavours $q=u,d,s$, since the heavy quark contributions (c,
b,...) could preferably be calculated perturbatively from the
intrinsic light quark ($u,d,s$) and gluon ($g$)
partonic-constituents of the nucleon.\\

Within the $\overline{MS}$ factorization scheme the NLO
contributions to $g_{1}^{p}(x,Q^{2})$ are finally given by
\cite{14}
\begin{equation}
g_{1}^{p}(x,Q^{2})= \frac{1}{2}\sum\limits_q e^2_q\bigl\{\delta
q(x,Q^2)+\delta \bar q(x,Q^2) +\frac{\alpha_s(Q^2)}{2\pi}[\delta
C_q\otimes (\delta q+\delta\bar q)+ 2\delta C_g\otimes\delta
g]\bigr\}\;,
\end{equation}\\
with the convolutions defined as
\begin{equation}
(C\otimes q)(x,Q^2)=\int\limits^1_x \frac{dy}{y} C(\frac{x}{y})
q(y,Q^2)\;.
\end{equation}

The polarized proton structure function, $xg_{1}^{p}$, in the LO
and NLO approximations can now be presented. The results for
$xg_{1}^{p}$ in the LO and NLO approximations as a function of $x$
and for some different values of $Q^2$ are depicted in Figs. 6,7.
The comparison between different models [21-23] and the available
experimental data
\cite{12, 13} have also been presented.\\

Since $\Delta (q+\overline{q})$ is the net number of right-handed
quarks of flavor $q$ inside a right-handed proton, it follows that
$\frac{1}{2}\Delta \Sigma $ is a measure of how much all quark
flavours contribute to the spin of the proton. Similarly, $\Delta
g$ represents the total gluonic contribution to the spin of the
nucleon. Using Eq.(2.20),  first moments of the polarized parton
distributions are calculated. These
results have been  shown in Table III.\\

Finally we can define, similarly to Eq.(2.20), the first moment of
$g_{1}^p$ (the Ellis-Jaffe sum rule) by \be \Gamma
_{1}^p(Q^{2})\equiv \int_{0}^{1}dxg_{1}^p(x,Q^{2})\;. \ee  The
results have also been given in Table III.

\subsection{Spin contribution of the valons to the proton}
The spin contribution of the valons to the proton can be analyzed
through the valon model. The contribution of various polarized
partons in a valon are calculable and by computing their first
moment, the spin of the proton can be computed. In the framework
of QCD the spin of the proton can be expressed in terms of the
first moment of the total quark and gluon helicity distributions
and their orbital angular momentum, i.e.
 \be
  \frac{1}{2}=\frac{1}{2}\Delta \Sigma
^{p}+\Delta g^{p}+L_{z}^{p}, \ee where $L_{z}^{p}$ is the total orbital
angular momentum of all the quarks and gluons.\\

Using Eq.(5.2)  the contributions of $\delta \Sigma(x,Q^2)$ and
$\delta g(x,Q^2)$ in the proton can be calculated.  These
contributions for just one $U$ valon in a proton can be extracted
as

\ba \delta \Sigma^U(x,Q^2)&=&\int_x^1  \delta
f^{S}(\frac{x}{y},Q^2)[\delta {\cal {W}}^{''}_{U}(y)\times
G_{U/p}(y)]\frac{dy}{y}\;,\nonumber \\
\delta g^U(x,Q^2)&=&\int_x^1  \delta
f^{gq}(\frac{x}{y},Q^2)[\delta {\cal {W}}^{'}_{U}(y)\times
G_{U/p}(y)]\frac{dy}{y}\;. \ea Using the above equations we can
arrive at their first moments as in below

\ba \Delta \Sigma^U&=&\int_0^1 \delta \Sigma^U(x,Q^2) dx\;,\nonumber \\
\Delta g^U&=&\int_0^1 \delta g^U(x,Q^2) dx\;. \ea

The results for a $D$-valon can be obtained in a similar way. The resulting total quark and gluon helicity for a
$U$-valon is \be
 \frac{1}{2}\Delta \Sigma ^{U}+\Delta g^{U}\;,
\ee and for $D$-valon  \be
 \frac{1}{2}\Delta \Sigma ^{D}+\Delta g^{D}\;.
\ee
 Since  each proton involves 2 $U$-valons and one $D$-valon, the total quark
 and gluon helicity for the proton is
\be
 \frac{1}{2}\Delta \Sigma ^{p}+\Delta g^{p}=2(\frac{1}{2}\Delta \Sigma ^{U}+\Delta
 g^{U})
+
 \frac{1}{2}\Delta \Sigma ^{D}+\Delta g^{D}\;.
\ee
 With the assumption that the polarized $U$ and $D$ valons have spin of $+\frac{1}{2}$ and
 $-\frac{1}{2}$  respectively, the
 determination of orbital angular momentum for each valon and
 finally for the proton, at different values of
 $Q^2$, can be obtained from Eq.(6.5).
  Our results at $Q^2=
 Q_0^2$ for the LO and NLO approximations are gathered in Table IV.
\\

\section{Conclusions}
We have used the valon model to describe  deep inelastic
scattering. The model bridges the gap between the bound state
problem and the scattering problem for hadrons. The valon
distribution can serve as a model for the solution to the bound
state problem. In contrast to this model, people usually use the
GLAP equations to evolve the parton distributions from an initial
value $Q_0$. The valon model which was first introduced by Hwa
[1-7,9], gives us a clear insight as to how to construct the
hadron structure from the parton distributions. According to this
model, a valon is defined as a cluster of valence quarks
accompanied by a cloud of sea quarks and gluons. It can be
considered as a bound state in which for instance a proton
consists of three valons, two U-valons  and one D-valon. This
model has been applied to the unpolarized case
and gives acceptable phenomenological results.\\

In this paper we extended the idea of the valon model to the
polarized case to describe the spin dependence of the hadron
structure functions. The polarized valon distribution is derived
from the unpolarized valon distribution. From the phenomenological
point view, we need to define weight functions to construct a
functional form for the polarized valon distribution. Due to the
difficulty which exists in deriving the spin contribution of the
sea quarks for all values $Q^2$ in the LO and at $Q^2=Q_0^2$ in
the NLO approximations, we improved the weight functions and
introduced two different types of them for the singlet and
non-singlet cases. In deriving the polarized valon distributions
some unknown parameters are present which should be determined by
fitting to the experimental data. Here we have used a method
similar to Ref.\cite{8}, to fit QCD predictions for the moments of
the $g_1^p$ structure function to the suitably constructed
Bernstein polynomial
averages of the E143 and SMC experimental data.\\

After calculating polarized valon distributions and all the parton
distributions in a valon, we calculated the polarized parton
density in a proton. The results were used to evaluate the spin
components of the proton. It turns out that our results for the
polarized structure functions are in good agreement with  all the
available experimental data of $g_{1}^p$ for the proton. Our
prediction for $\Gamma_{1}^p$, the first moment of $g_1^p$, is
also in good
agreement with the  experimental data.\\

 We hope to report
in future on application  of the improved valon model in
describing more complicated hadron structure functions. We also
hope to be able to consider the symmetry breaking of polarized sea
quarks in our new calculations.

\section{Acknowledgments}
We are grateful to R.C.Hwa for giving us his useful and
constructive comments. We would like to thank C.J.Maxwell and
M.M.Sheikh Jabbari for reading and correcting the manuscript of
this paper and for useful discussions. A.N.K is grateful to CERN
for their hospitality whilst he visited there and could amend this
paper. We acknowledge the Institute for Studies in Theoretical
Physics and Mathematics (IPM) for financially supporting this
project. A.N.K. thanks Semnan university and S.A.T thanks  Persian
Gulf university for partial financial support of this project.

\newpage

\textbf{Table Caption}\\

\begin{center}
\begin{tabular}{|c|c|c|c|}
\hline\hline $Q^2\;GeV^2$&$\cal{A}$&$\cal{B}$&$\cal{C}$\\ \hline
     3      &  0.274  &  0.876  &  1.212  \\
     5      &  0.339  &  0.911  &  1.724  \\
    10      &  0.342  &  0.842  &  1.932  \\ \hline\hline
\end{tabular}
\end{center}
\textbf{Table I} Numerical values of fitting
${\cal{A}},{\cal{B}},{\cal{C}}$ parameters in Eq.(4.5).
\\
\\

\textbf{LO}
\\
\begin{tabular}{ccccccc}
\hline\hline $j$ & \multicolumn{1}{|c}{$N_{j}$} & $\alpha _{j}$ &
$\beta _{j}$ & $\gamma _{j}$ & $\eta _{j}$ \\ \hline
$U$ & \multicolumn{1}{|c}{0.002} & -2.3789 & -1.7518 & 11.0804 & -1.4629 \\
$D$ & \multicolumn{1}{|c}{-0.005} & -1.5465 & -1.8776 & 8.5042 & -0.8608 \\
\hline\hline &  &  &  &  &  &\\ \hline\hline ${\cal{A}}_{0}$ &
${\cal{A}}_{1}$ &
${\cal{A}}_{2}$ & ${\cal{A}}_{3}$ & ${\cal{A}}_{4}$ & ${\cal{A}}_{5}$ & $\Lambda_{\overline{MS}}\; (MeV)$ \\
\hline 0.00043 & 0.2954 &
-6.9134 & 30.9851 & -39.7383 & 16.4605 & 203 \\ \hline\hline \\
\end{tabular}
\\
\\
\textbf{NLO}
\\
\begin{tabular}{ccccccc}
\hline\hline $j$ & \multicolumn{1}{|c}{$N_{j}$} & $\alpha _{j}$ &
$\beta _{j}$ & $\gamma _{j}$ & $\eta _{j}$ \\ \hline
$U$ & \multicolumn{1}{|c}{0.0038} & -2.1501 & -0.8859 & 10.6537 & -0.1548 \\
$D$ & \multicolumn{1}{|c}{-0.0046} & -1.5859 & -1.5835 & 9.6205 & -0.8410 \\
\hline\hline &  &  &  &  &  \\ \hline\hline ${\cal{A}}_{0}$ &
${\cal{A}}_{1}$ & ${\cal{A}}_{2}$ & ${\cal{A}}_{3}$ &
${\cal{A}}_{4}$ & ${\cal{A}}_{5}$ & $\Lambda_{\overline{MS}}\; (MeV)$\\
\hline -0.0025 & -3.1148
& 15.8114 & -21.1500 & 10.5025 & -0.9162 & 235 \\ \hline\hline \\ \\
\end{tabular}
\\
\\
\textbf{Table II} Numerical values of fitting parameters for the
best fit of Fig. 2 in the LO and NLO approximations.
\\
\\
\begin{center}
\begin{tabular}{ccccccc}
\hline\hline
$Q^{2}(GeV^{2})$ & $\Delta u_{v}$ & $\Delta d_{v}$ & $\Delta \Sigma $ & $%
\Delta \overline{q}$ & $\Delta g$ & $\Gamma_1^p$ \\ \hline
1  & 0.8807 & -0.3328 & 0.1786 & -0.0615 & 0.5480 & 0.1214 \\
3  & 0.8774 & -0.3315 & 0.1724 & -0.0622 & 0.7774 & 0.1242 \\
5 & 0.8765 & -0.3311 & 0.1706 & -0.0624 & 0.8737 & 0.1249 \\
10 & 0.8754 & -0.33308 & 0.1688 & -0.0626 & 0.9986 & 0.1258 \\
\hline\hline
\end{tabular}
\end{center}
\textbf{Table III} The first moments of polarized parton
distributions, $\Delta u_{v}$, $\Delta d_{v}$, $\Delta \Sigma $,
$\Delta \overline{q}$, $\Delta g$ and $\Gamma_1^p$  in the  NLO
approximation for some value of $Q^{2}$.
\\
\\
\\
\\
\\
\\
\begin{center}
\begin{tabular}{|c|c|c|c|c|}
\hline\hline
$i$ & $\frac{1}{2}\Delta \Sigma ^{i}$ & $\Delta g^{i}$ & $L_{z}^{i}$ & $%
\frac{1}{2}\Delta \Sigma ^{i}+\Delta g^{i}+L_{z}^{i}$ \\ \hline
$U$ & $0.091$ & $0.440$ & $-0.031$ & $0.5$ \\ \hline $D$ &
$-0.077$ & $-0.332$ & $-0.091$ & $-0.5$ \\ \hline $P$ & $0.105$ &
$0.548$ & $-0.153$ & $0.5$ \\ \hline\hline
\end{tabular}
\end{center}
\hspace{7 cm} \textbf{(a)}
\\
\\
\begin{center}
\begin{tabular}{|c|c|c|c|c|}
\hline\hline
$i$ & $\frac{1}{2}\Delta \Sigma ^{i}$ & $\Delta g^{i}$ & $L_{z}^{i}$ & $%
\frac{1}{2}\Delta \Sigma ^{i}+\Delta g^{i}+L_{z}^{i}$ \\ \hline
$U$ & $0.039$ & $0.440$ & $0.020$ & $0.5$ \\ \hline $D$ & $0.011$
& $-0.332$ & $-0.178$ & $-0.5$ \\ \hline $P$ & $0.089$ & $0.548$ &
$-0.137$ & $0.5$ \\ \hline\hline
\end{tabular}
\end{center}
\hspace{7 cm} \textbf{(b)}\\

 \textbf{Table IV} Spin contribution
of  valons and proton in the LO ($\bf a$) and the  NLO ($\bf b$)
approximations at $Q^2=Q_0^2$.
\\
\\
\newpage
\textbf{Figure Caption}\\
\\
\\
{\bf{Fig. 1}}-The contribution of moments $NS$, $S$ and $gq$ from
Eqs.(3.4, 3.6) as a function of $n$ at $Q^2=3\; GeV^2$.
Dashed-dotted line and  solid line indicate respectively  the  LO
and  the NLO approximations.\\ \\
{\bf{Fig. 2}}-{Fit to $xg_1^p$ using the Bernstein averages.
Dashed-dotted line and  solid line indicate respectively  the  LO
and  the NLO approximations.}\\ \\
{\bf{Fig. 3}}-{ The plots of total polarized valon distributions
in the proton, $\sum _{j}^{}y\delta G_{j/p}$ ,  for non-singlet
and singlet polarized valon distributions as a function of $y$
where $j$ runs over $U,U,D$ valons.}\\ \\
{\bf{Fig. 4}}-{Polarized parton distributions in the proton at
$Q^2$=3 GeV$^2$ as a function of $x$ in the LO approximation. The
solid line is our model, dashed line is the AAC model
(ISET=1)[21], dashed-dotted line is the BB model (ISET=1)[22] and
long-dashed line is the GRSV model (ISET=3)[23].}\\ \\
{\bf{Fig. 5}}-{Polarized parton distributions in the proton at
$Q^2$=3 GeV$^2$ as a function of $x$ in the NLO approximation. The
solid line is our model, dashed line is the AAC model
(ISET=3)[21], dashed-dotted line is the BB model (ISET=3)[22] and
long-dashed line is the GRSV model (ISET=1)[23].}\\ \\
{\bf{Fig. 6}}-{Polarized proton structure function $xg_{1}^{p}$ as
a function of $x$ which is compared with the experimental data
from Ref.[12,13] for different $Q^2$ values in the LO
approximation. The solid line is our model, dashed line is the AAC
model (ISET=1)[21], dashed-dotted line is the BB model
(ISET=1)[22] and dashed-dotted-dotted line is the GRSV model
(ISET=3)[23].}\\ \\
{\bf{Fig. 7}}-{Polarized proton structure function $xg_{1}^{p}$ as
a function of $x$ which is compared with the experimental data
from Ref.[12,13] for different $Q^2$ values in the NLO
approximation. The solid line is our model, dashed line is the AAC
model (ISET=3)[21], dashed-dotted line is the BB model
(ISET=3)[22] and dashed-dotted-dotted line is the GRSV model
(ISET=1)[23].}

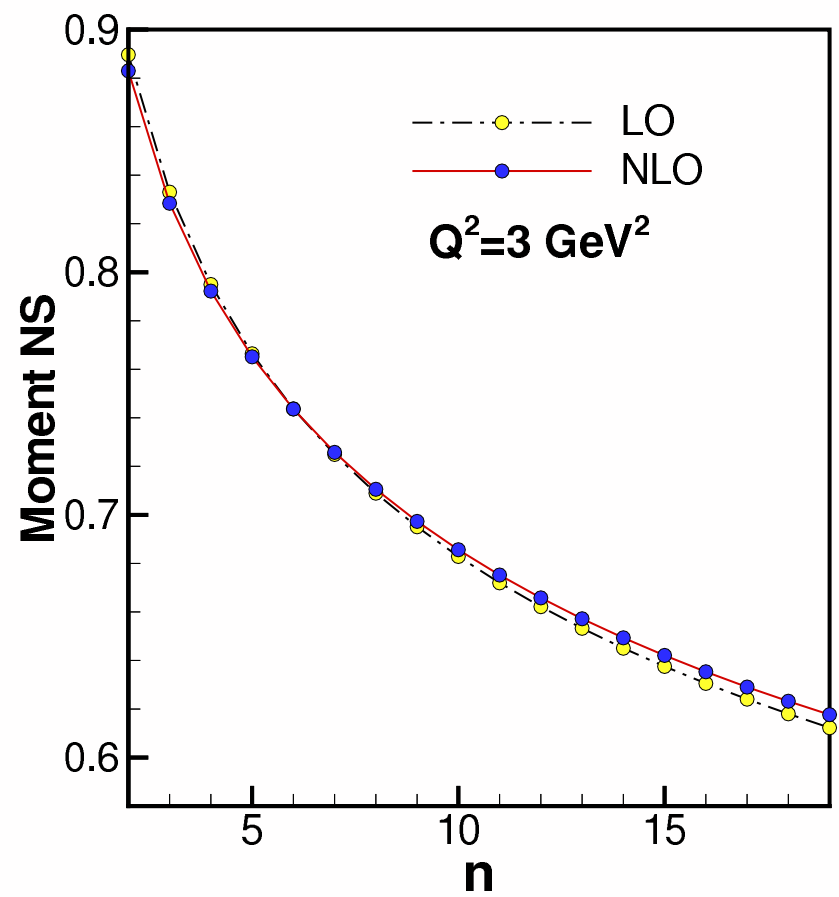
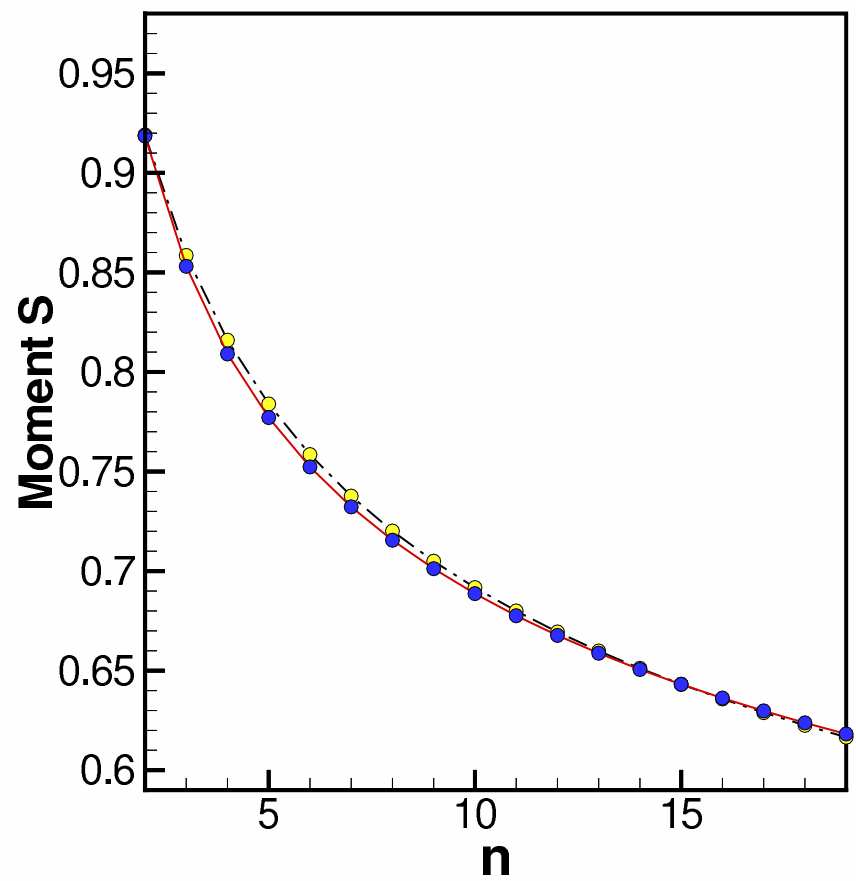
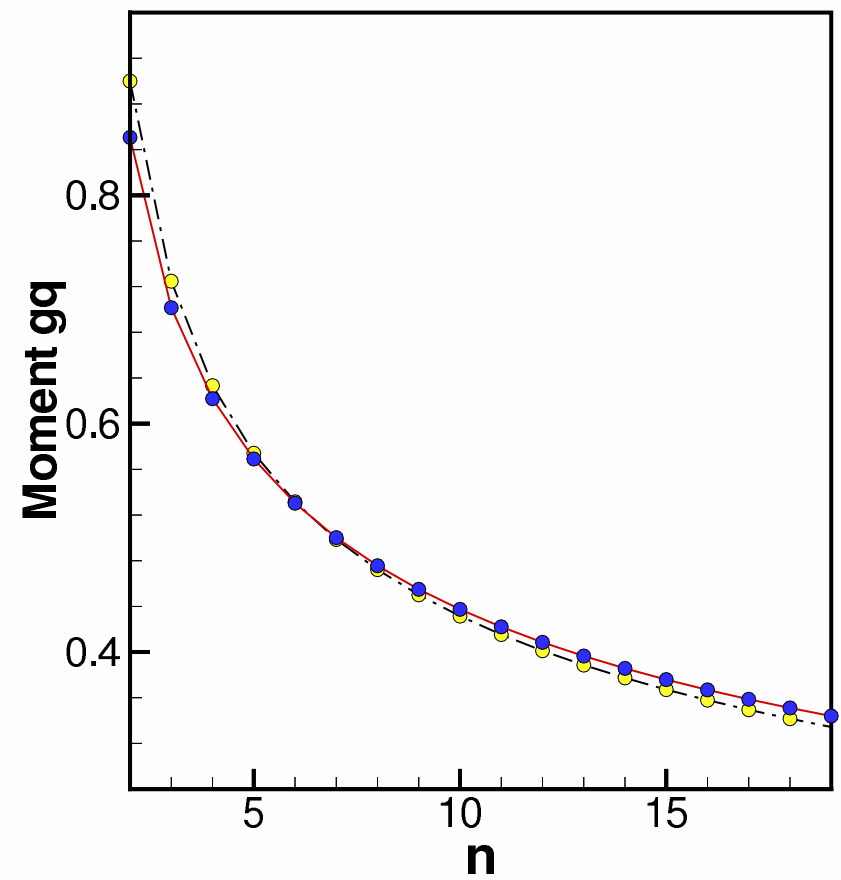
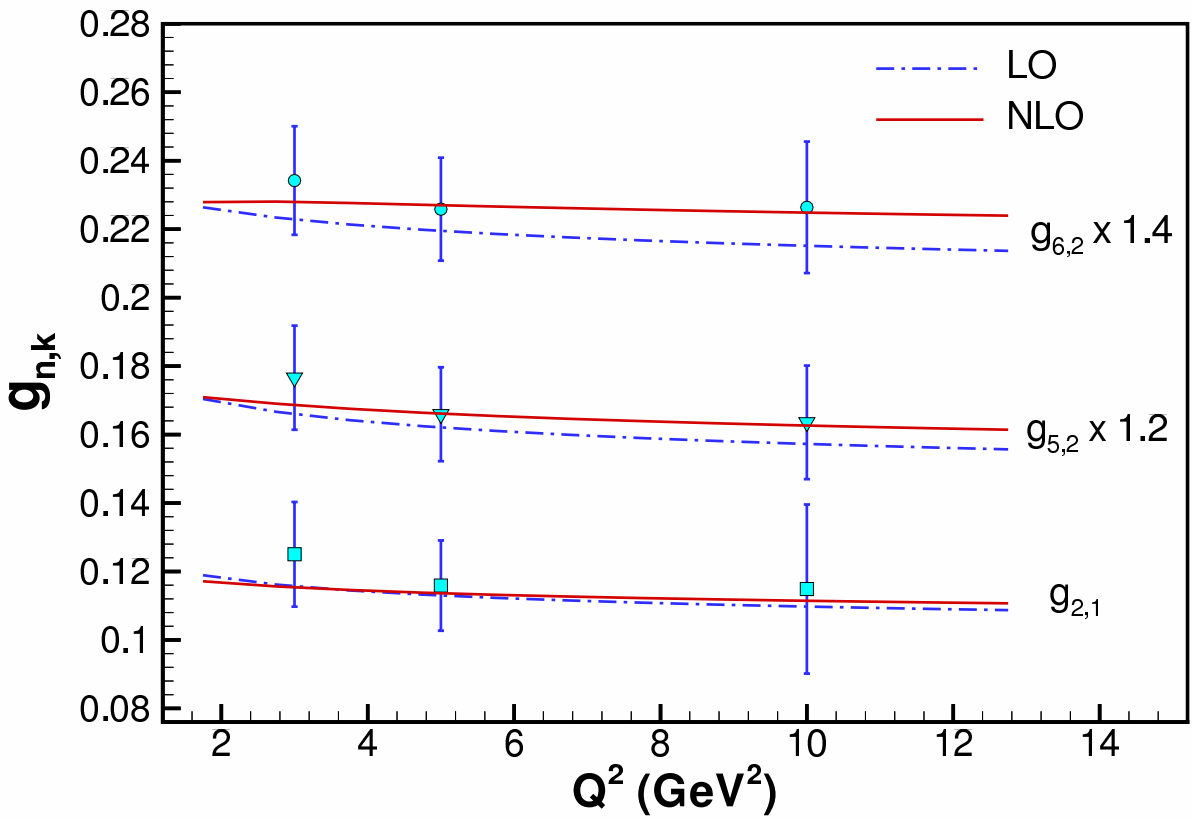
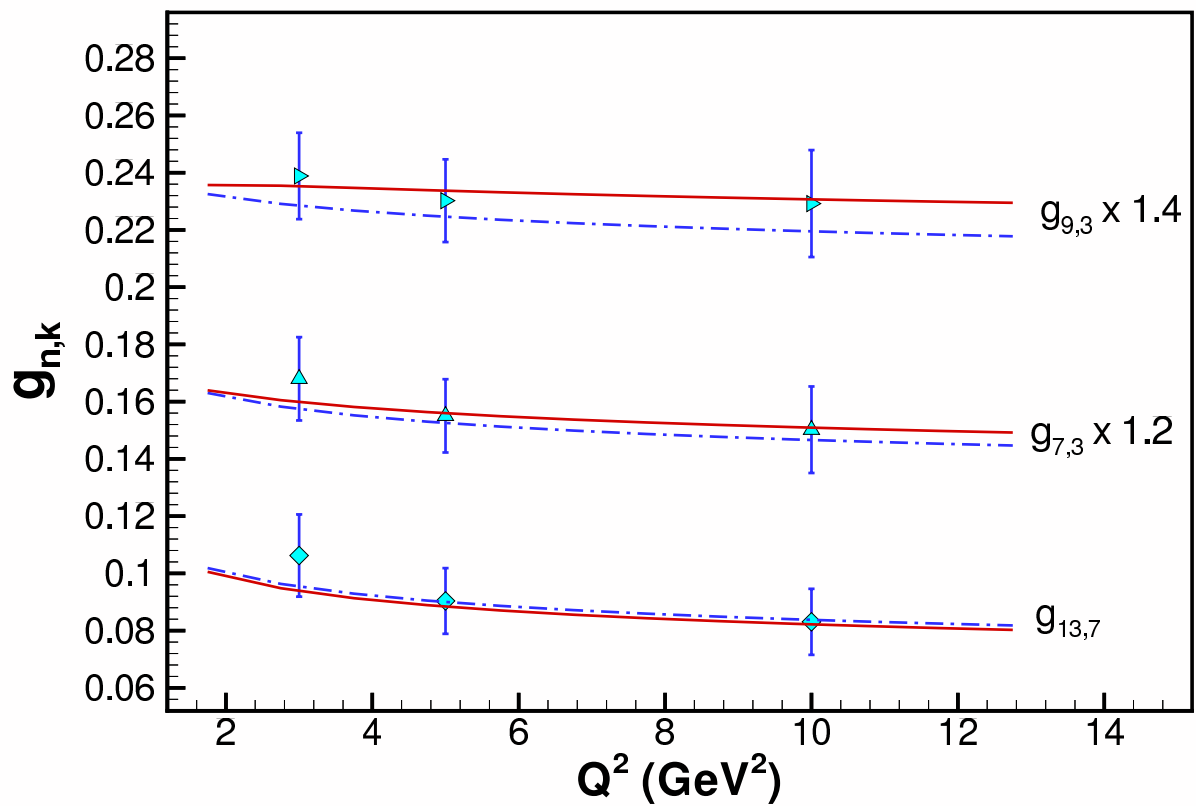
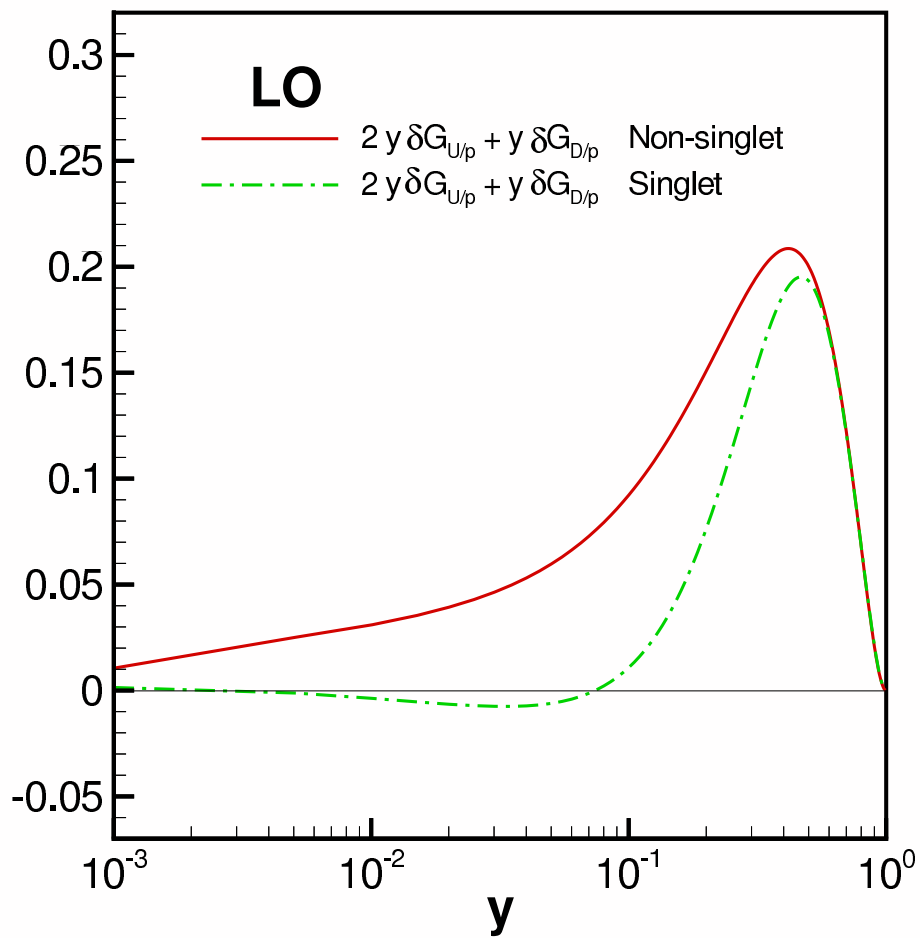
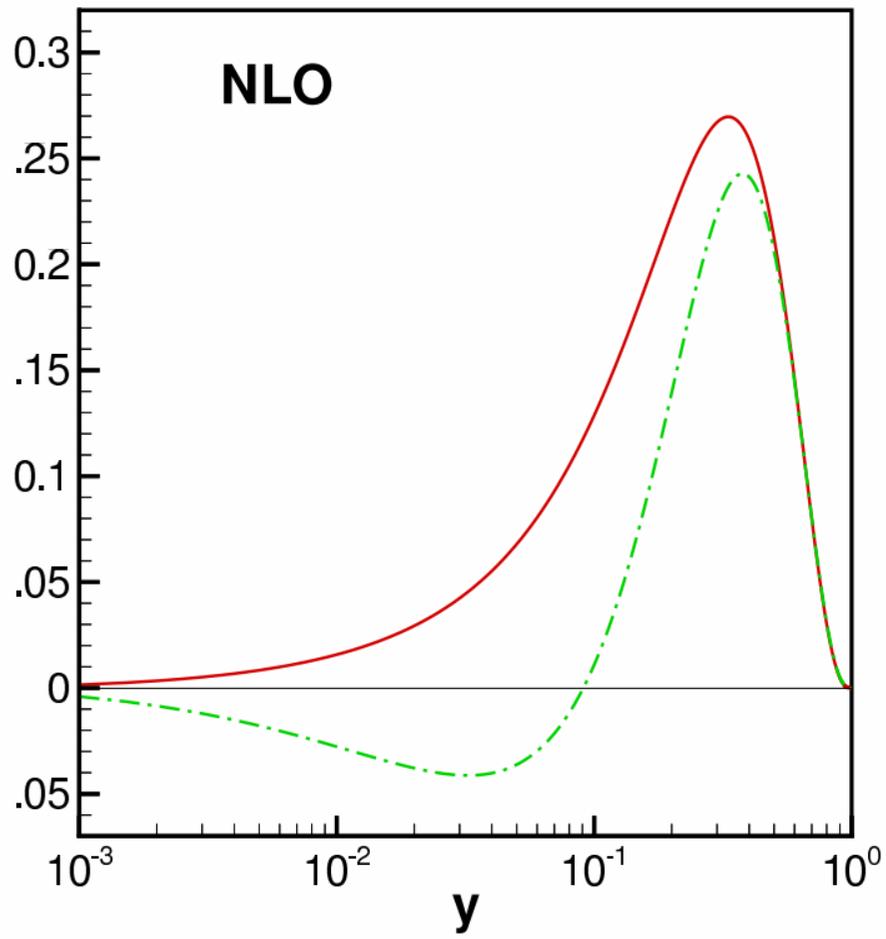
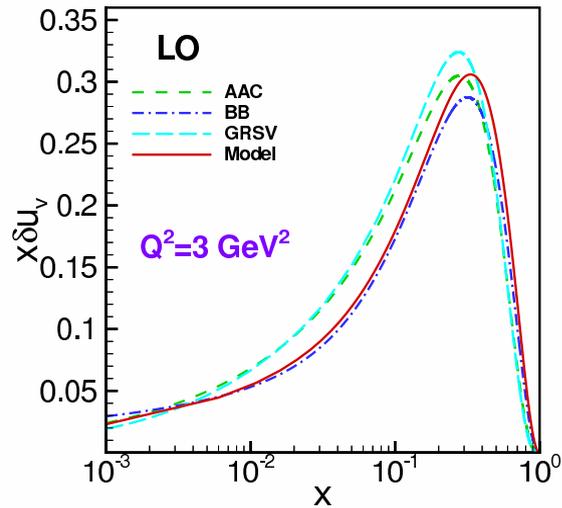
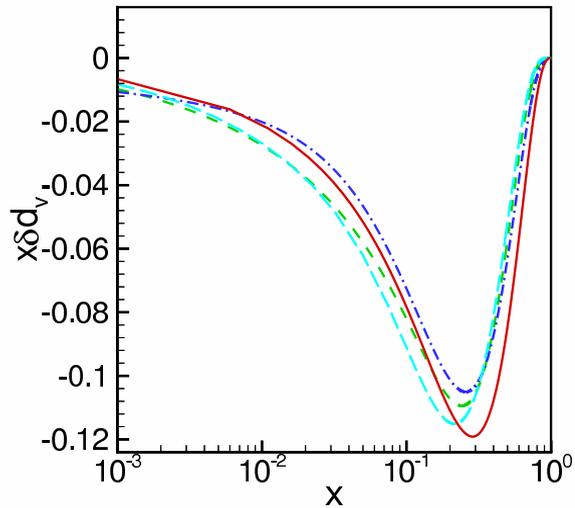
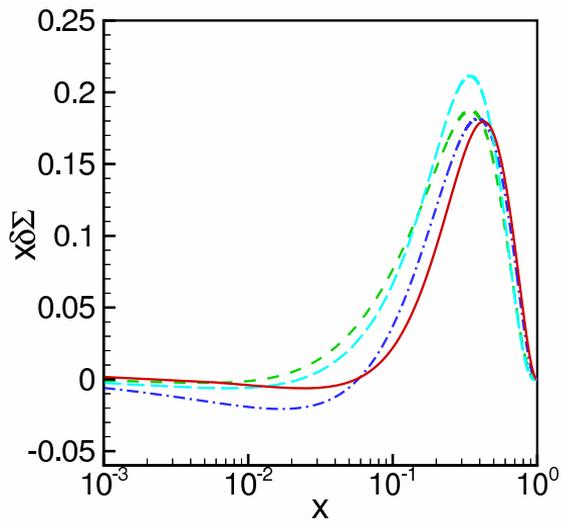
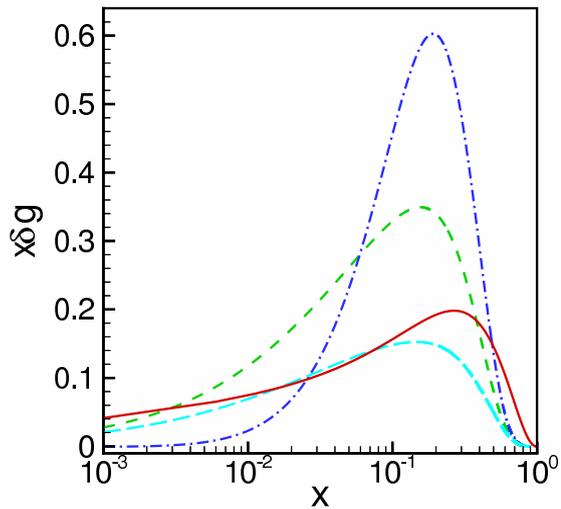
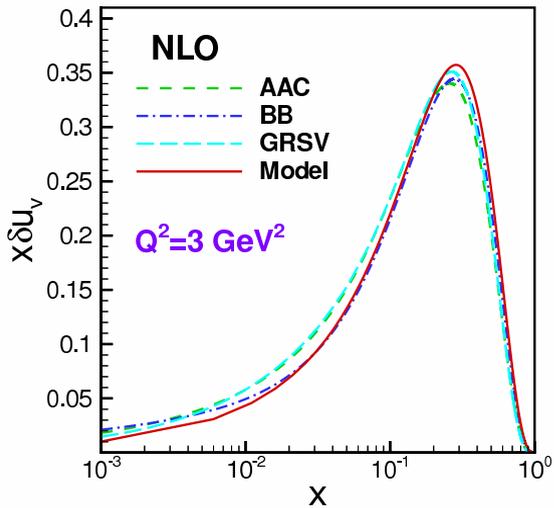
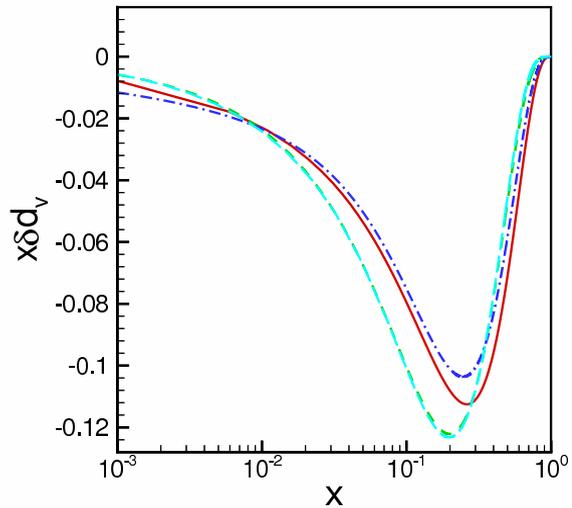
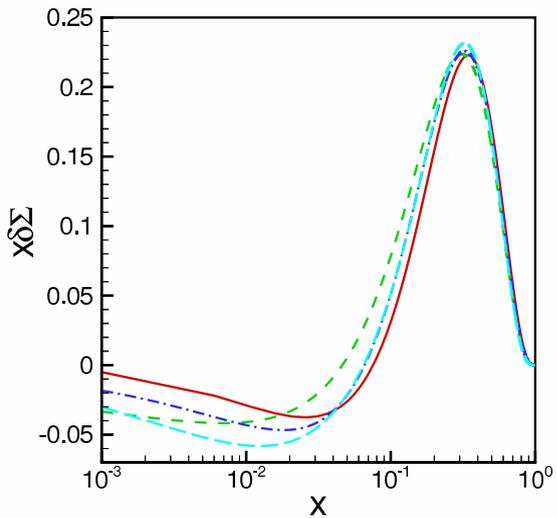
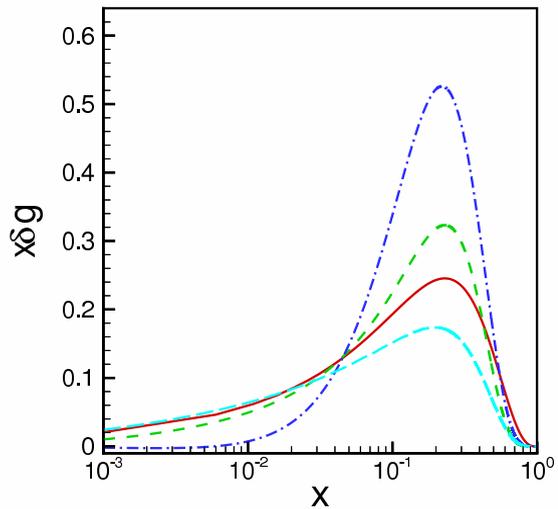
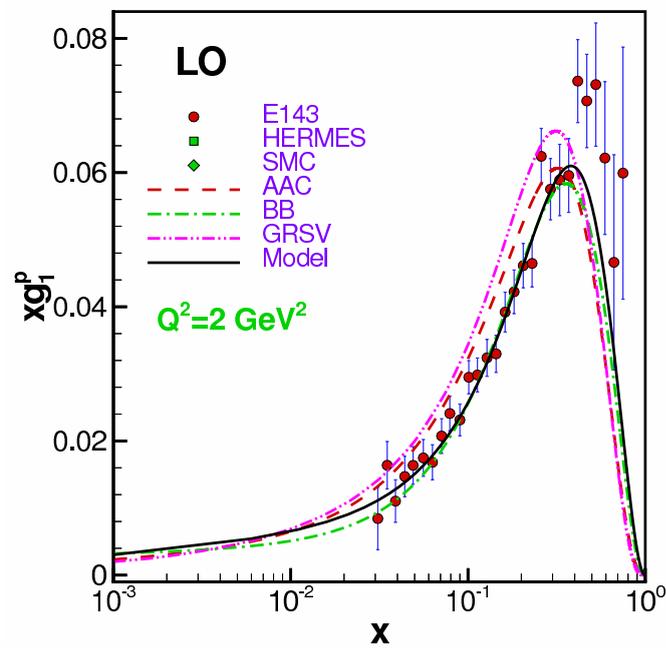
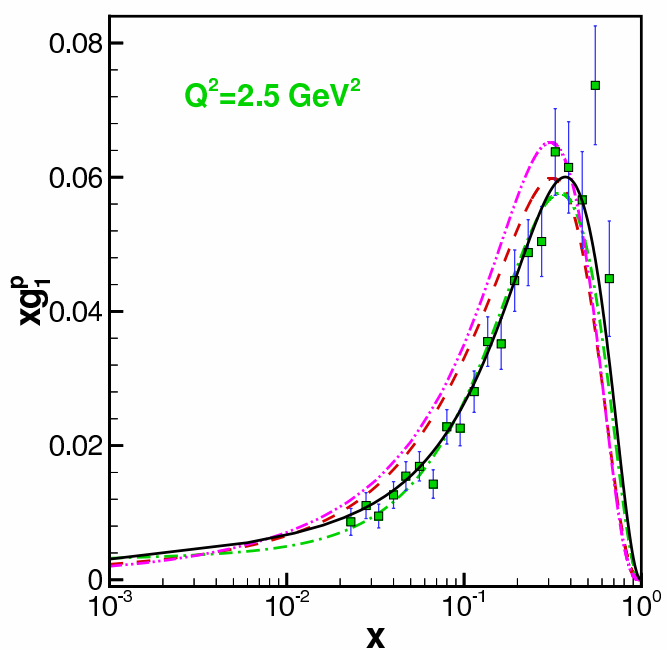
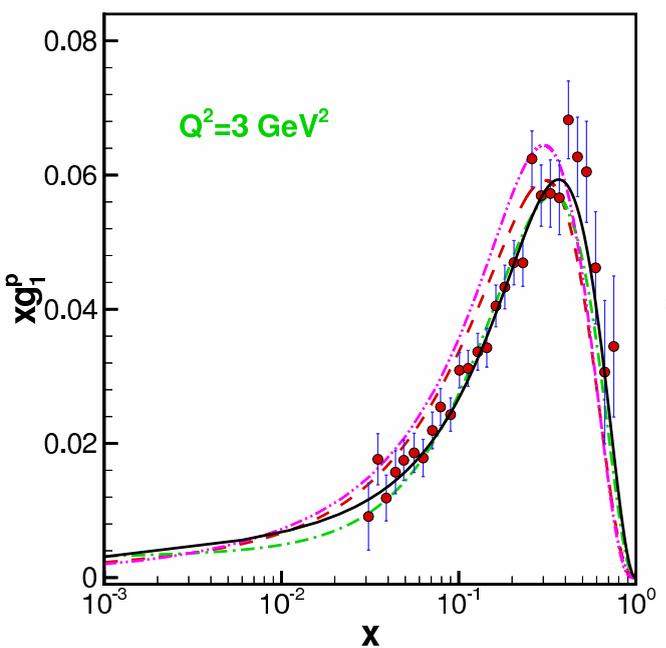
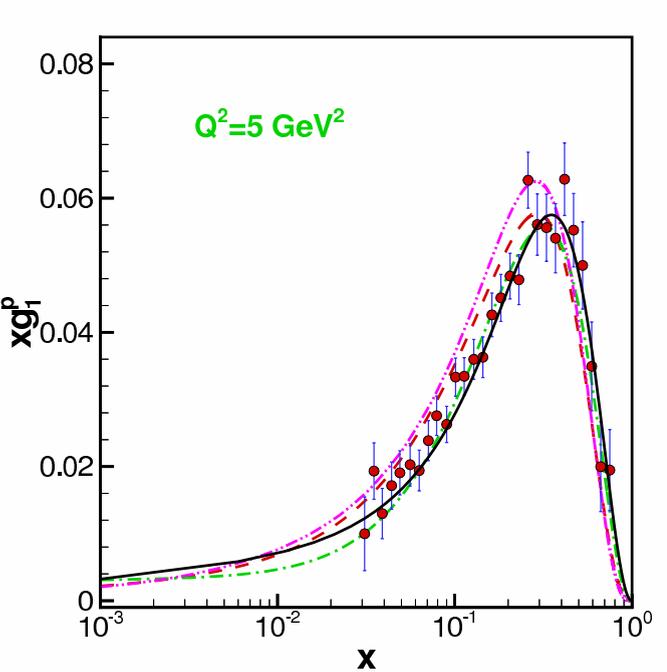
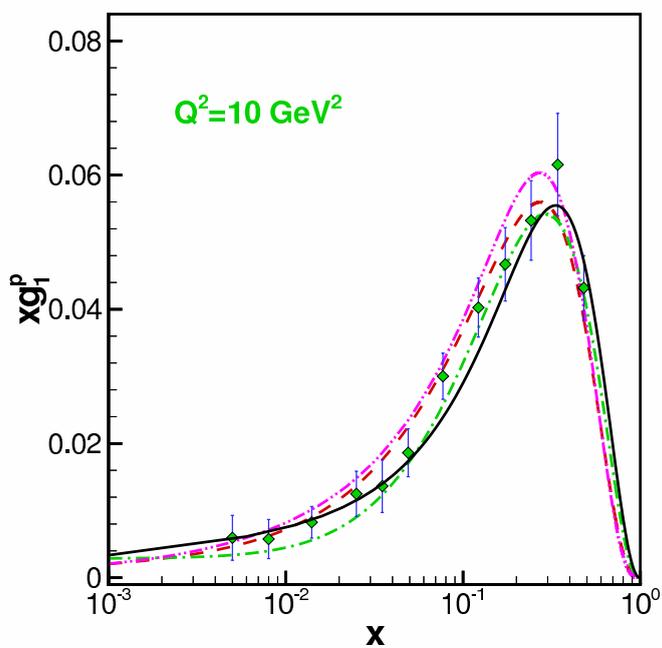
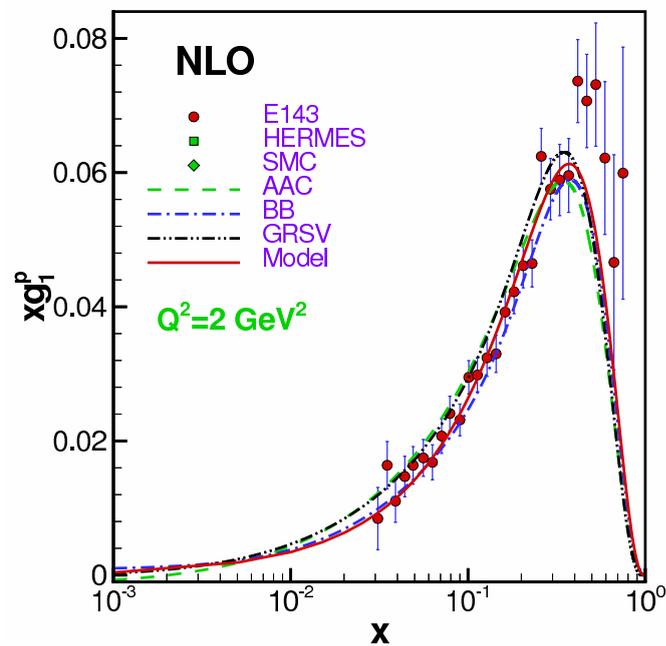
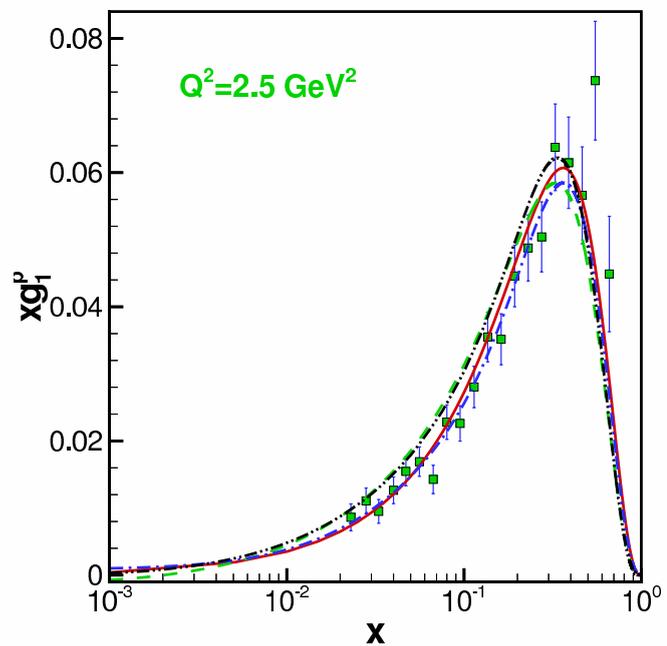
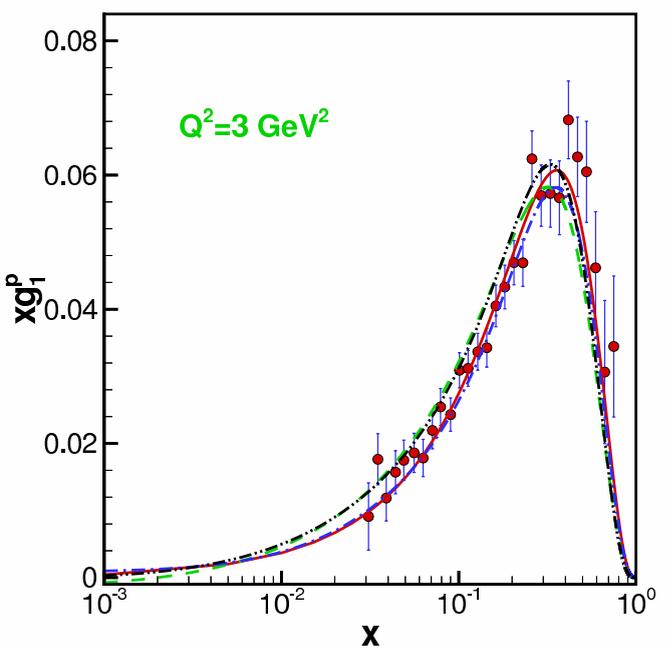
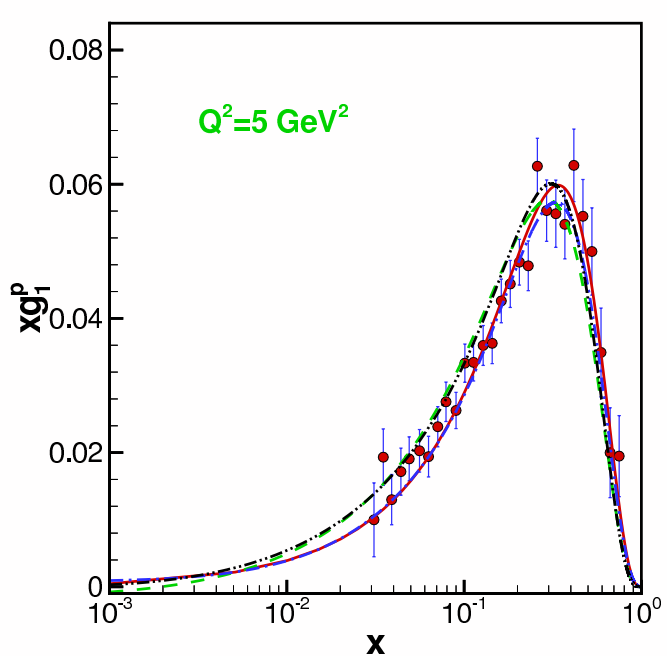
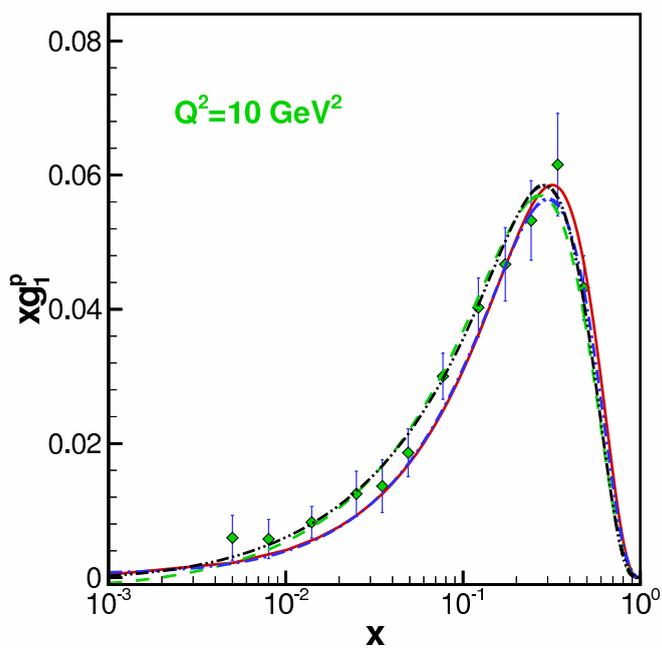
\end{document}